\documentclass[12pt]{article}
\pdfoutput=1
\usepackage{amssymb,cite,graphicx}
\usepackage{slashed}
\usepackage{amsmath,bm,bbm}
\usepackage{amsfonts}
\usepackage[titletoc,title]{appendix}
\usepackage[small]{caption}
\usepackage[margin=1in]{geometry}
\usepackage[multiple]{footmisc}
\usepackage{mathtools}
\usepackage{slashed}
\usepackage[nottoc]{tocbibind}
\usepackage{xcolor}

\definecolor{mediumblue}{rgb}{0,0,0.8}
\usepackage{hyperref}
\hypersetup{
  linktocpage=true,
  colorlinks=true,
  citecolor=mediumblue,
  filecolor=mediumblue,
  linkcolor=mediumblue,
  urlcolor=mediumblue,
}

\graphicspath{{./}{./figures/}}
\numberwithin{equation}{section}
\allowdisplaybreaks%

\newcommand{\be}{\begin{equation}}
\newcommand{\ee}{\end{equation}}
\newcommand{\bea}{\begin{eqnarray}}
\newcommand{\eea}{\end{eqnarray}}
\begin{document}

\begin{titlepage}
  \begin{flushright}
 CERN-TH-2020-001
  \end{flushright}
  \medskip

  \begin{center}
    {\Large\bf\boldmath
    Dark matter self-interactions from spin-2 mediators
    }
\vspace{1.5cm}

    {\bf Yoo-Jin Kang$^{1,\star}$ and Hyun Min Lee$^{1,2,\dagger}$ }

    {\it $^1$Department of Physics, Chung-Ang University, Seoul
      06974, Korea}\\[0.2cm]
          {\it  $^2$CERN, Theory department, 1211 Geneva 23, Switzerland}\\[0.2cm]
  \end{center}

  \bigskip

  \begin{abstract}
    \noindent
   We propose a new mechanism for rendering  dark matter self-interacting in the presence of a massive spin-2 mediator.
The derived Yukawa-type potential for dark matter is independent of the spins of dark matter in the leading order of the momentum expansion, so are the resulting non-perturbative effects for the dark matter self-scattering. We find that both the Born cross section and relatively mild resonance effects assist to make the self-scattering cross section velocity-dependent. We discuss how to evade the current indirect bounds on dark matter annihilations and show that the model is marginally compatible with perturbative unitarity in the ghost-free realization of the massive spin-2 particle.

  \end{abstract}

\vspace{3.5cm}
  \begin{flushleft}
  $^\star$Email: yoojinkang91@gmail.com \\
    $^\dagger$Email: hminlee@cau.ac.kr 
  \end{flushleft}
\end{titlepage}

\section{Introduction}

%intro
There are plenty of indirect evidences for dark matter (DM) such as galaxy rotation velocities, gravitational lensing, large scale structures, Cosmic Microwave Background (CMB) anisotropies, etc. 
It has been assumed that dark matter is collision-less, so there is no or little self-interaction between dark matter particles. Weakly Interacting Massive Particles (WIMPs) have been a well motivated candidate for dark matter with negligible self-interaction and weak interactions with known particles in the Standard Model but they have been challenged by strong bounds from direct detection experiments \cite{xenon1t}. Any single evidence for dark matter beyond the gravitational interactions would provide an important guideline for pinning down the particle physics nature of dark matter.

%small-scale
There has been a tension between $N$-body simulations and observed rotation velocities in galaxies. The former favors the cuspy profile of dark matter density distribution at galaxies but the latter shows the cored profiles. This is known as the small-scale problem \cite{smallscale,smallscale2}, which is related to another problem such as too-big-to-fail problem. Self-Interacting Dark Matter (SIDM) has been suggested to solve those small-scale problems via the large self-scattering cross section with $\sigma_{\rm self}/m_{\rm DM}=0.1-10\,{\rm cm^2/g}$ \cite{Yu}. 
Although baryonic effects, if included in the $N$-body simulations, could ease or eliminate the tension \cite{baryon}, it is worthwhile to investigate the particle physics models for rendering DM self-interactions velocity-dependent to be consistent with the bounds from galaxy clusters \cite{bullet} and look for the observable signatures.

%spin-2 mediator
In this article, we propose a novel mechanism for self-interacting dark matter of arbitrary spin by exchanging a massive spin-2 mediator between dark matter particles. The spin-2 mediator couples to dark matter through the energy-momentum tensor \cite{GMDM,DD,GLDM}, giving rise to the effective Yukawa-type potential between dark matter particles. In this framework, we compute the momentum transfer cross section for DM self-scattering in the Born limit and include the non-perturbative effects for the same process in the presence of a light spin-2 mediator.
We also show how the DM self-scattering cross section is velocity-dependent in order to satisfy the bounds from galaxy clusters. We also discuss the consistency of large self-interactions with indirect bounds on dark matter annihilations and perturbative unitarity in the presence of non-linear spin-2 couplings. 

The readers can refer to a companion paper of the same authors  \cite{SIDM-spin2} dealing with the effective theory for dark matter self-interactions with a massive spin-2 mediator, which includes a complete discussion in the momentum expansions of the dark matter self-interactions in the effective field theory and contains the next-to-leading order terms and spin-dependent interactions beyond the leading terms that we focus on in this work. Therefore, the companion paper in Ref.~\cite{SIDM-spin2} is complementary to our current paper.

%paper

\section{Dark matter potential from spin-2 mediators}

We introduce the couplings of a massive spin-2 mediator ${\cal G}_{\mu\nu}$ with mass $m_G$ to the SM particles and dark matter with mass $m_{\rm DM}$ (which is a real scalar $S$, a Dirac fermion $\chi$ or a real vector $X$), through the energy-momentum tensor, as follows \cite{GMDM},
\bea
{\cal L}_{\rm int}= -\frac{c_{\rm SM}}{\Lambda} {\cal G}^{\mu\nu} T^{\rm SM}_{\mu\nu} -\frac{c_{\rm DM}}{\Lambda} {\cal G}^{\mu\nu} T^{\rm DM}_{\mu\nu}.
\eea
Then, the tree-level scattering amplitude for the self-scattering of dark matter through the spin-2 mediator is 
\bea
{\cal M}=-\frac{c^2_{\rm DM}  }{\Lambda^2} \frac{i}{q^2-m^2_G}\,T^{\rm DM}_{\mu\nu}(q){\cal P}^{\mu\nu,\alpha\beta}(q) T^{\rm DM}_{\alpha\beta}(-q)  \label{ampl}
\eea
where $q$ is the 4-momentum transfer between dark matter particles and the tensor structure for the massive spin-2 propagator is given by
\bea
{\cal P}_{\mu\nu,\alpha\beta}(q)=\frac{1}{2}\Big(G_{\mu\alpha}G_{\nu\beta}+G_{\nu\alpha} G_{\mu\beta}- \frac{2}{3} G_{\mu\nu} G_{\alpha\beta}\Big)
\eea
with
\bea
G_{\mu\nu}\equiv \eta_{\mu\nu}- \frac{q_\mu q_\nu}{m^2_G}. 
\eea
Here, we note that the energy-momentum tensor for dark matter, $T^{\rm DM}_{\mu\nu}$, depends not only on the 4-momentum transfer but also on the dark matter momenta, although it is not explicitly shown.
The tensor $P_{\mu\nu,\alpha\beta}$ satisfies traceless and transverse conditions for on-shell spin-2 mediator, such as $\eta^{\alpha\beta} P_{\mu\nu,\alpha\beta}(q)=0$ and $q^\alpha P_{\mu\nu,\alpha\beta}(q)=0$ \cite{GMDM}.
A similar approach was taken for computing the DM-nucleon scattering amplitudes in the effective field theory with a massive spin-2 mediator and dark matter \cite{DD,GLDM}.

The conservation law $q_\mu T^{\mu\nu}=0$ is satisfied for $q$ being the 4-momentum of the massive spin-2 mediator mediated between on-shell dark matter particles, so we can replace $G_{\mu\nu}$ in the scattering amplitude (\ref{ampl}) by $\eta_{\mu\nu}$. 
For instance, the energy-momentum tensor for fermion dark matter $\chi$ is given by
\bea
T^\chi_{\mu\nu}= -\frac{1}{4} {\bar u}_\chi(k_2)\Big(\gamma_\mu (k_{1\nu}+k_{2\nu})+\gamma_\nu (k_{1\mu}+k_{2\mu})-2\eta_{\mu\nu}(\slashed{k}_1+\slashed{k}_2-2m_\chi) \Big)u_\chi(k_1)
\eea
where the fermion DM is incoming into the vertex with momentum $k_1$ and is outgoing from the vertex with momentum $k_2$. Then, we can show explicitly that $q^\mu T^\chi_{\mu\nu}=0$ with $q^\mu=k^\mu_1-k^\mu_2$ being the momentum of the massive spin-2 mediator by using the equation of motion for fermion dark matter.  Similarly, the general energy-momentum tensors containing other spins of dark matter and/or the SM particles follow the same conservation law, $q^\mu T_{\mu\nu}=0$.

As a consequence, the self-scattering amplitude for dark matter in eq.~(\ref{ampl}) is divided into trace and traceless parts of energy-momentum tensor, as follows,
\bea
{\cal M}=-\frac{c^2_{\rm DM} }{2\Lambda^2} \frac{i}{q^2-m^2_G}\, \bigg(2 T^{\rm DM}_{\mu\nu} T^{{\rm DM},\mu\nu} -\frac{2}{3}(T^{\rm DM})^2 \bigg).
\eea
As a consequence, in the non-relativistic limit of dark matter and $m_G\lesssim m_{\rm DM}$, we find that the effective potential for dark matter is approximated to be Yukawa-type, up to $(m_G/m_{\rm DM})^2$ corrections \cite{SIDM-spin2}, independent of the spins of dark matter,  as follows,
\bea
V_{{\rm eff}} \simeq - \frac{A_{\rm DM}}{4\pi r}\, e^{-m_G r} \label{Yuk}
\eea
with
\bea
A_{\rm DM} =\frac{2c^2_{\rm DM} m^2_{\rm DM}}{3 \Lambda^2}. \label{fine}
\eea
Therefore, the effective self-coupling $A_{\rm DM}$ of dark matter is determined by the DM mass and the gravitational coupling to the spin-2 mediator.  We note that both spin-independent and spin-dependent effective field potentials for dark matter with the massive spin-2 mediator were derived in Ref.~\cite{SIDM-spin2}.

\section{Spin-2 mediators and dark matter self-interactions}

We first discuss the Born cross section for dark matter self-scattering and derive the Yukawa type potential for dark matter self-scattering in the non-perturbative regime. Then, we show the parameter space for self-scattering cross section in the Hulth\'en potential approximation and comment on the potential problem from dark matter annihilations and solutions.

\subsection{Born approximations for self-scattering}

The momentum transfer cross section for DM self-scattering \cite{haibo,kai} is given by
\bea
\sigma_T = 2\pi \int^1_{-1}\frac{d\sigma}{d\Omega}\, \Big(1-|\cos\theta|\Big) d\cos\theta. \label{ssx}
\eea
We first consider the Born regime with  $A_{\rm DM} m_{\rm DM}/(4\pi m_G)\lesssim 1$ and take the limit of a small dark matter velocity with $m_{\rm DM}v\lesssim m_G$ where $v$ is the relative velocity of dark matter.
Then,  the momentum transfer cross sections for DM self-scattering 
are given in the order of scalar, fermion and vector dark matter, as follows,
\bea
\sigma_{S, T}^{\rm Born}&\simeq &{A^2_S\over 4\pi m^2_G v^2 }{\ln\Big(1+{m_S^2 v^2 \over m_G^2}\Big)\over \Big(1 + {m_S^2v^2\over 2m_G^2}\Big)^3}, \label{born1}  \\
\sigma^{\rm Born}_{\chi,T}&\simeq&{A^2_\chi\ \over 8\pi m^2_\chi v^4}\bigg[ \bigg(1+ {2m_\chi^2 m_G^4 v^2 \over(m_\chi^2 v^2 + 2m_G^2)^3} \bigg) \ln\Big(1+{m_\chi^2 v^2 \over m_G^2} \Big) \nonumber \\
&&-{m_\chi^2 v^2 \over m_\chi^2 v^2 +m_G^2} \bigg], \label{born2} \\
\sigma_{X,T}^{\rm Born}&\simeq&{A^2_X \over 12\pi m^2_G v^2}{(32-56r_X+27r_X^2)\over (4-r_X)^2}\,{\ln\Big(1+{m_X^2 v^2 \over m_G^2}\Big)\over \Big(1 + {m_X^2v^2\over 2m_G^2}\Big)^3},  \label{born3}
\eea
with $A_{\rm DM}$ being defined in eq.~(\ref{fine}) for ${\rm DM}=S,\chi, X$ and $r_X=(m_G/m_X)^2$.
These approximate results in the Born limit are used to compare with the full results in the later discussion in Fig.~\ref{vs}.
In the limit of a vanishing DM velocity, we can approximate  eqs.~(\ref{born1}), (\ref{born2}) and (\ref{born3}) further, as follows,
\bea
\sigma_{S, T}^{\rm Born}&\simeq&{A^2_S m^2_S\over 4\pi m^4_G} \bigg(1-\frac{2m^2_S v^2}{m_G^2} \bigg), \\
\sigma^{\rm Born}_{\chi,T}&\simeq &{3A^2_\chi m^2_\chi \over 32\pi m^4_G} \bigg(1-\frac{14 m_\chi^2v^2}{9 m^2_G} \bigg),\\
\sigma_{X,T}^{\rm Born}&\simeq&{A^2_X m^2_X \over 12\pi m^4_G }{(32-56r_X+27r_X^2)\over (4-r_X)^2} \,\bigg(1-\frac{2m^2_Xv^2}{m^2_G}\bigg),
\eea
which differ from the total self-scattering cross sections at the leading order in Ref.~\cite{GLDM} by $1/2, 3/4$ and $1/2$ factors for scalar, fermion and dark matter cases, respectively, due to the fact that the momentum transfer is not averaged over in the latter case.

\subsection{Bethe-Salpeter equation with spin-2 mediator}

In the non-perturbative regime with $A_{\rm DM} m_{\rm DM}/(4\pi m_G)\gtrsim 1$, Sommerfeld and/or bound-state effects become more important.  In the Coulomb limit with a small dark matter velocity, we need to resum the ladder diagrams with the massive spin-2 mediator for the self-scattering of dark matter in the Feynman diagram approach as in the cases with light spin-0 or spin-1 mediators \cite{haibo,kai}, resulting in a Schr\"odinger-like equation with the Yukawa-type potential for dark matter given in eq.~(\ref{Yuk}).  
For consistency, we will also show  in the next section that the spin-2 mediator coupling does not exceed the unitarity bound for dark matter annihilation processes.

Before going into a further discussion on the self-scattering and Sommerfeld effects for dark matter, we discuss the resummation of the ladder diagrams in the case of the massive spin-2 mediator in more detail. As illustration, we consider the elastic self-scattering process for scalar dark matter, $S(p)+S(k)\rightarrow S(p')+S(k')$. We find that the non-perturbative four-point function $\Gamma(p,k;p',k')$ for the scattering process with the spin-2 mediator exchanges satisfies a recursive relation  \cite{cassel} , as follows,
\bea
i\Gamma(p,k;p',k') &=& i\Gamma(p,k;p',k') -\int\frac{d^4s}{(2\pi)^4} \, {\tilde\Gamma}(p,k;p+k-s,s)  \nonumber \\
&\times&  G(s) G(p+k-s) \Gamma(p+k-s,s;p',k') \label{recursive}
\eea
where $G(s)$ is the propagator for scalar dark matter and ${\tilde \Gamma}(p,k;p',k')$ is the tree-level four-point amplitude, given by
\bea
{\tilde \Gamma}(p,k;p',k')&=&-\frac{2c^2_S}{\Lambda^2} \frac{1}{(p-p')^2-m^2_G} \bigg[ A(p,k) A(p'.k')  \nonumber \\
&&+ A(p,k')A(p',k) -\frac{2}{3} A(p,p') A(k,k') \bigg] \label{treeampl}
\eea
with
\bea
A(p,k)\equiv p\cdot k-\frac{(p\cdot q)(k\cdot q)}{m^2_G},\qquad q=p-p'.
\eea

The resummation of ladder diagrams is needed to capture Sommerfeld effects at a small momentum transfer between dark matter particles due to $t$-channel poles. The typical momentum transfer for dark matter self-scattering is $q\sim m_{\rm DM} v$.
Since the momentum integration is dominated by small loop momenta for energy-momentum conservation in the non-relativistic self-scattering of dark matter,
 the energy transfer becomes $\omega=p_0-p'_0\approx 0$, thus the scattering process is instantaneous, and we can approximate the above tree-level point amplitude in eq.~(\ref{treeampl}) \cite{SIDM-spin2} to 
\bea
{\tilde \Gamma}(p,k;p',k')&\approx& \frac{8 c^2_S m^4_S}{3\Lambda^2}\, \frac{1}{{\vec q}^2+m^2_G}\,\bigg[1+\frac{3}{2} (v^\perp)^2+\frac{3}{8}(v^\perp)^4 \nonumber \\
&&\quad +\frac{{\vec q}^2}{4 m^2_S}\Big(1+\frac{3}{2} (v^\perp)^2\Big)-\frac{{\vec q}^4}{4 m^2_S m^2_G} \Big(1-\frac{m^2_G}{4m^2_S}\Big) \bigg] \label{tapprox}
\eea
where ${\vec q}={\vec p}-{\vec p}'$ is the momentum transfer between dark matter particles, ${\vec v}^\perp\cdot {\vec q}=0$ and $(v^\perp)^2=v^2-\frac{{\vec q}^2}{m^2_S}$ with $v$ being the relative velocity between dark matter particles.

Therefore, defining the Bethe-Salpeter(BS) wave function in momentum space for dark matter in the following,
\bea
{\tilde\psi}_{\rm BS} = \int \frac{dP_0}{2\pi} \, \chi (P,Q)
\eea
where $P=\frac{1}{2} (p+k)$, $Q=\frac{1}{2}(p-k)$, and 
\bea
\chi(p,k) \equiv G(p) G(k) \Gamma(p,k;p',k'),
\eea
and using eq.~(\ref{recursive}) with eq.~(\ref{tapprox}) while ignoring the perturbative contributions, 
we obtain the BS equation for the wave function in position space as
\bea
-\frac{1}{m_S} \nabla^2 \psi_{\rm BS} ({\vec x} )+V({\vec x}) \psi_{\rm BS} ({\vec x}) =E\, \psi_{\rm BS} ({\vec x})
\eea
with the effective potential being given by
\bea
V({\vec x}) &=& -\frac{1}{4m^2_S} \int \frac{d^3{\vec q}}{(2\pi)^3} e^{i{\vec q}\cdot {\vec x}}\,\cdot{\tilde \Gamma}(p,k;p',k')  \nonumber \\
&=&-\frac{c^2_S m^2_S}{6\pi \Lambda^2 r}\, e^{-m_G r}\bigg[1+\frac{3}{2} (v^\perp)^2+\frac{3}{8}(v^\perp)^4 -\frac{m^2_G}{2m^2_S}\Big(1+\frac{3}{4} (v^\perp)^2\Big)+\frac{m^4_G}{16 m^4_S}  \bigg] . \label{Yuk2}
\eea
As a result, the correction terms coming from nonzero velocity $v^\perp$ and momentum transfer are suppressed as far as $|v^\perp|\ll 1$ and $m_G\ll m_S$. Therefore, even higher order terms in the momentum expansion of the tree-level amplitude with the massive spin-2 exchange give rise to suppressed contributions to the effective potential.

We remark on the validity of the momentum expansion for the effective potential. The effective potential  in eq.~(\ref{Yuk2}) is given by the infinite momentum integral for the momentum transfer. However, since the resummation of ladder diagrams is dominated by a small momentum transfer near the spin-2 particle mass, we can truncate the effective potential up to finite terms in the effective field theory. An explicit cutoff or regularization on the momentum transfer was introduced in Ref.~\cite{tanedo} in order to treat the higher order terms in the momentum transfer. But, in our case, as far as we keep the momentum transfer small for the self-scattering of dark matter in the effective theory, higher momentum contributions are sub-dominant for the computation of the effective potential, so the small momentum expansion of the tree-level self-scattering amplitude  as in eq.~(\ref{tapprox}) is justified.
The above result  in eq.~(\ref{Yuk2}) is consistent with eq.~(\ref{Yuk}) in the limit of $m_G\ll m_S$ and $(v^\perp)^2\ll 1$. 
The same discussion holds for fermion or vector dark matter as well, apart from the spin-dependent parts of the effective potential \cite{SIDM-spin2}. 
As a result, we have shown that  the non-perturbative amplitude for dark matter self-scattering can be computed consistently in the case of the massive spin-2 mediator even with the non-renormalizable interactions.

\subsection{Loop corrections due to spin-2 mediators}

In this subsection, we also comment on the loop corrections of the massive spin-2 mediator to the self-scattering of dark matter. Concretely, we consider the one-loop corrections to the self-scattering amplitude for scalar dark matter with two massive spin-2 particles exchanged.  
Then, as summarized in Appendix A, in the non-relativistic limit for dark matter, we can approximate the $t$-channel scattering amplitude to
\bea
i\Gamma_{\rm loop}= i\Gamma_{\rm div} + i\Gamma_{\rm finite}
\eea
where $\Gamma_{\rm div}$ is the divergent part, in dimensional regularization, given by
\bea
i\Gamma_{\rm div}= \frac{c^4_S m^6_S}{24\pi^2\Lambda^4 m^4_G}\, (10m^2_G + 7m^2_S)\, \cdot\frac{1}{\epsilon},
\eea
and $\Gamma_{\rm finite}$ is the finite part, obtained in the limit of $\xi\equiv m^2_S/m^2_G\gg 1$ as
\bea
i\Gamma_{\rm finite} \approx \frac{2 c^4_S m^6_S}{ 9\pi \Lambda^4 m^2_G} \,\sqrt{\xi}.
\eea
We also obtained the same results from the $u$-channel diagrams as for the $t$-channel diagrams in the non-relativistic limit.
Here, the divergent part $i\Gamma_{\rm div}$ can be cancelled by the renormalization of the quartic self-coupling for scalar dark matter, $ -\frac{1}{4}\lambda_S S^4$. 

Now we also discuss the finite part of the loop corrections to the self-scattering amplitude. 
We first recall the tree-level amplitude for small momentum and momentum transfer for scalar dark matter from eq.~(\ref{tapprox}) as
\bea
i{\tilde\Gamma} \approx  \frac{8c^2_S m^4_S}{3\Lambda^2 m^2_G}.
\eea
As a result, in the limit of $\xi\equiv m^2_S/m^2_G\gg 1$, we obtain the ratio of the finite part of the one-loop amplitude to the tree-level amplitude for scalar dark matter, as follows,
\bea
\frac{\Gamma_{\rm finite}}{\tilde\Gamma} \approx \frac{c^2_S m^2_S}{12\pi \Lambda^2}\,\sqrt{\xi}=\frac{1}{8} A_S \sqrt{\xi}. 
\eea
Therefore, for $\xi\gg 1$ and $A_S\lesssim 1$, we get $\Gamma_{\rm finite}\gtrsim {\tilde\Gamma}$, for which the perturbative expansion would break down as in the cases for spin-0 or spin-1 mediators \cite{reece}, so we need to resum the ladder diagrams following the Bethe-Salpeter formalism as discussed in the previous subsection. In the non-perturbative regime for the one-loop amplitude, we obtain the condition, $m_G\lesssim c^2_S m^3_S/(12\pi \Lambda^2)$, which is similar to the non-perturbative condition for the bound-state formation of dark matter, as will be discussed in the next subsection, 

As shown in Appendix A, we also checked that the leading velocity-dependent corrections in the one-loop amplitude are divergent but they can be cancelled by higher dimensional counter terms for scalar dark matter, such as $c_6(\partial_\mu S\partial^\mu S) S^2$, etc. The finite velocity-dependent corrections at one loop can be ignored as far as $4 v^2 \xi\lesssim 1$ with $v$ being the relative velocity between dark matter particles. This is the case for dark matter in galaxies and galaxy clusters.

Although the concrete discussion on the loop corrections was made for the case for scalar dark matter for simplicity, the similar results could be also obtained to the cases for fermion or vector dark matter.

\subsection{Hulth\'en potential approximation}

In the non-perturbative regime with $A_{\rm DM} m_{\rm DM}/(4\pi m_G)\gtrsim 1$,  the self-scattering cross section for dark matter can be also enhanced by non-perturbative and resonance effects.
We take the Hulth\'en potential approximation for the Yukawa-type effective potential (\ref{Yuk}) for dark matter, 
with $V_H= -\frac{A_{\rm DM}}{4\pi } \frac{\delta e^{-\delta r}}{1-e^{-\delta r}}$, with $\delta=\frac{\pi^2}{6} m_G$.
Then, for the $s$-wave dominance, the general result for the non-perturbative self-scattering cross section is given by
\bea
\sigma^{\rm Hulthen}_T\simeq \frac{4\pi \sin^2\delta_0}{k^2}. \label{selfH}
\eea 
where the phase shift for the $s$-wave is given by
\bea
\delta_0={\rm arg} \left(\frac{i\Gamma(\lambda_++\lambda_--2)}{\Gamma(\lambda_+)\Gamma(\lambda_-)} \right) \label{phaseshift}
\eea
with
\bea
\lambda_\pm=1 + \frac{ik}{\delta} \pm \sqrt{\eta^2-\frac{k^2}{\delta^2}}, \qquad \eta=\sqrt{ \frac{A_{\rm DM} m_{\rm DM}}{4\pi \delta}.}\label{lam-eta}
\eea
At the pole of the gamma function at $\lambda_-=-n$ in the phase shift, with $n$ being non-negative integer, the self-scattering cross section is enhanced by $\sigma_{\rm self}\propto 1/v^2$. 
In this case, dark matter can form an $s$-wave bound state for $\omega x=n^2$ for a positive integer $n$ \cite{cassel}, leading to the resonance condition for the spin-2 mediator mass,
\bea
m_G=\frac{3}{2\pi^3 n^2}\, A_{\rm DM} m_{\rm DM}=\frac{c^2_{\rm DM}}{\pi^3 n^2}\, \frac{ m^3_{\rm DM}}{\Lambda^2}. \label{resonance}
\eea
This is an intriguing relation between the masses for the spin-2 mediator and dark matter  and the strength of the spin-2 mediator coupling. We note that the similar condition as above was also inferred from the explicit calculations of the one-loop corrections to the self-scattering amplitude for dark matter in the previous subsection.

\begin{figure}[tbp]
\centering
\includegraphics[width=.40\textwidth]{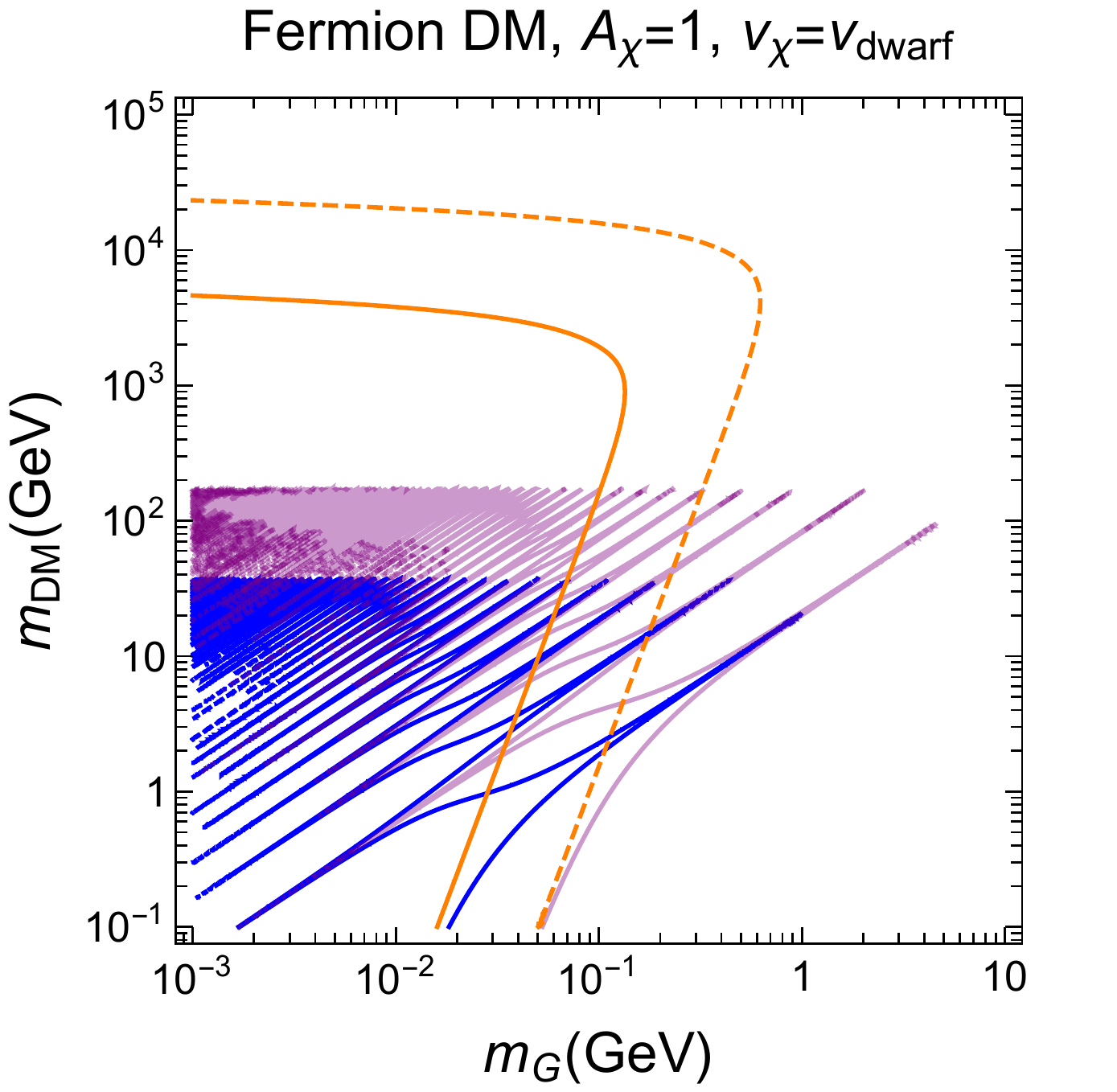}\,\,
\includegraphics[width=.40\textwidth]{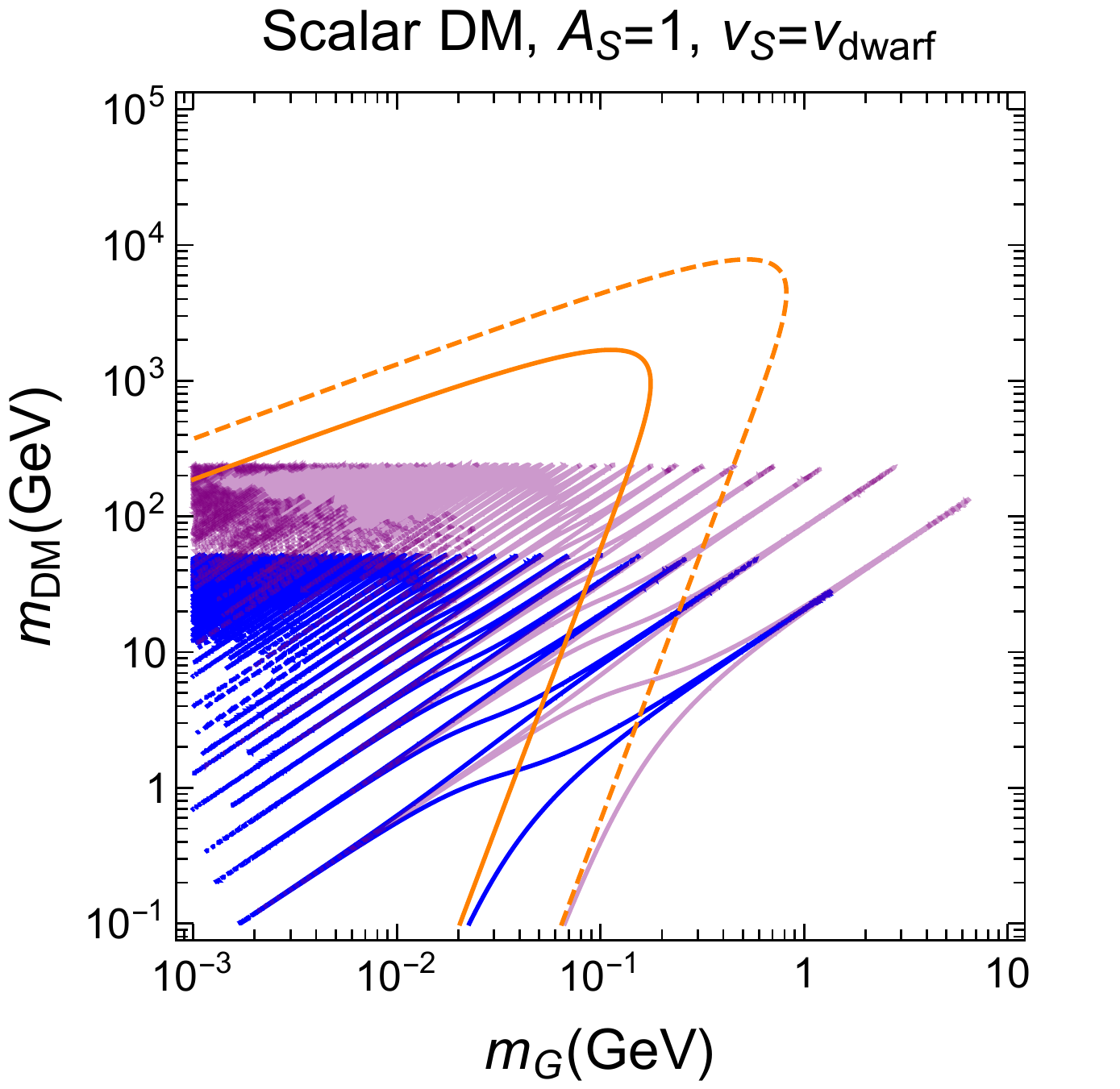}
  \caption{Contours of dark matter self-scattering cross sections in $m_G$ vs $m_{\rm DM}$, depending on the spins of dark matter, $s=1/2,0$ on left and right. We have chosen $A_{\rm DM}=1$ and the DM velocity to the one at dwarf galaxies, $v_{\rm dwarf}=10^{-4}$.   The orange dashed and solid lines are the results for the Born cross section, whereas the purple and blue lines are those for the non-perturbative cross section. We took $\sigma_T/m_{\rm DM}=0.1,10\,{\rm cm^2/g}$. The case with $s=1$ shows the similar result as for the case with $s=0$.  }
  \label{mg}
\end{figure}

In matching the non-perturbative results to the Born approximations given in eqs.~(\ref{born1}), (\ref{born2}) and (\ref{born3}), we make replacements for the self-scattering cross sections, depending on the spins of dark matter, as follows,
\bea
\sigma_{S,T}&\simeq&\frac{\sigma^{\rm Hulthen}_T}{(\psi^{(2)}(1))^2(6/\pi^2)^4},  \label{exact-S} \\
\sigma_{\chi,T} &\simeq &\frac{3}{8}  \frac{\sigma^{\rm Hulthen}_T}{(\psi^{(2)}(1))^2(6/\pi^2)^4}, \label{exact-F}  \\
\sigma_{X,T} &\simeq &\frac{(32-56r_X+27 r^2_X)}{3(4-r_X)^2} \cdot \frac{\sigma^{\rm Hulthen}_T}{(\psi^{(2)}(1))^2(6/\pi^2)^4}\,\cdot\label{exact-V}
\eea
where $\sigma^{\rm Hulthen}_T$ and $\eta$  in eqs.~(\ref{selfH}) and (\ref{lam-eta}) are given by those with $A_{\rm DM}$ being replaced by $A_S, A_\chi$ and $A_X$ in order).
For our analysis on the dark matter self-scattering, we use the above analytic results.

In Fig.~\ref{mg}, we depicted the contours in the parameter space for $m_G$ vs $m_{\rm DM}$ for the DM self-scattering cross section divided by the DM mass. We have fixed the DM velocity to $v_{\rm dwarf}=10^{-4} c$  at dwarf galaxies, the effective fine structure constant to $A_{\rm DM}=1$, and the contours are shown for $\sigma_T/m_{\rm DM}=0.1, 10\,{\rm cm^2/g}$. The orange dashed and solid lines indicate the results with the Born cross section. On the other hand, the results with the non-perturbative cross section are shown in purple and blue lines. The cases for  fermion and scalar dark matter are shown on left and in the panel. The case for vector dark matter is similar to the case for scalar dark matter, so we don't show it in Fig.~\ref{mg}.
We found that the DM masses up to $200\,{\rm GeV}$ and the spin-2 mediator masses up to $6\,{\rm GeV}$ are required to get the self-scattering cross section for  solving the small-scale problems. We find that fermion dark matter is distinguishable from scalar or vector dark matter, due to the difference in the Born cross section. This is because the particle-particle and particle-anti-particle scattering processes coexist in the case of fermion dark matter, unlike in the other cases.

\begin{figure}[tbp]
  \begin{center}
    \includegraphics[height=0.32\textwidth]{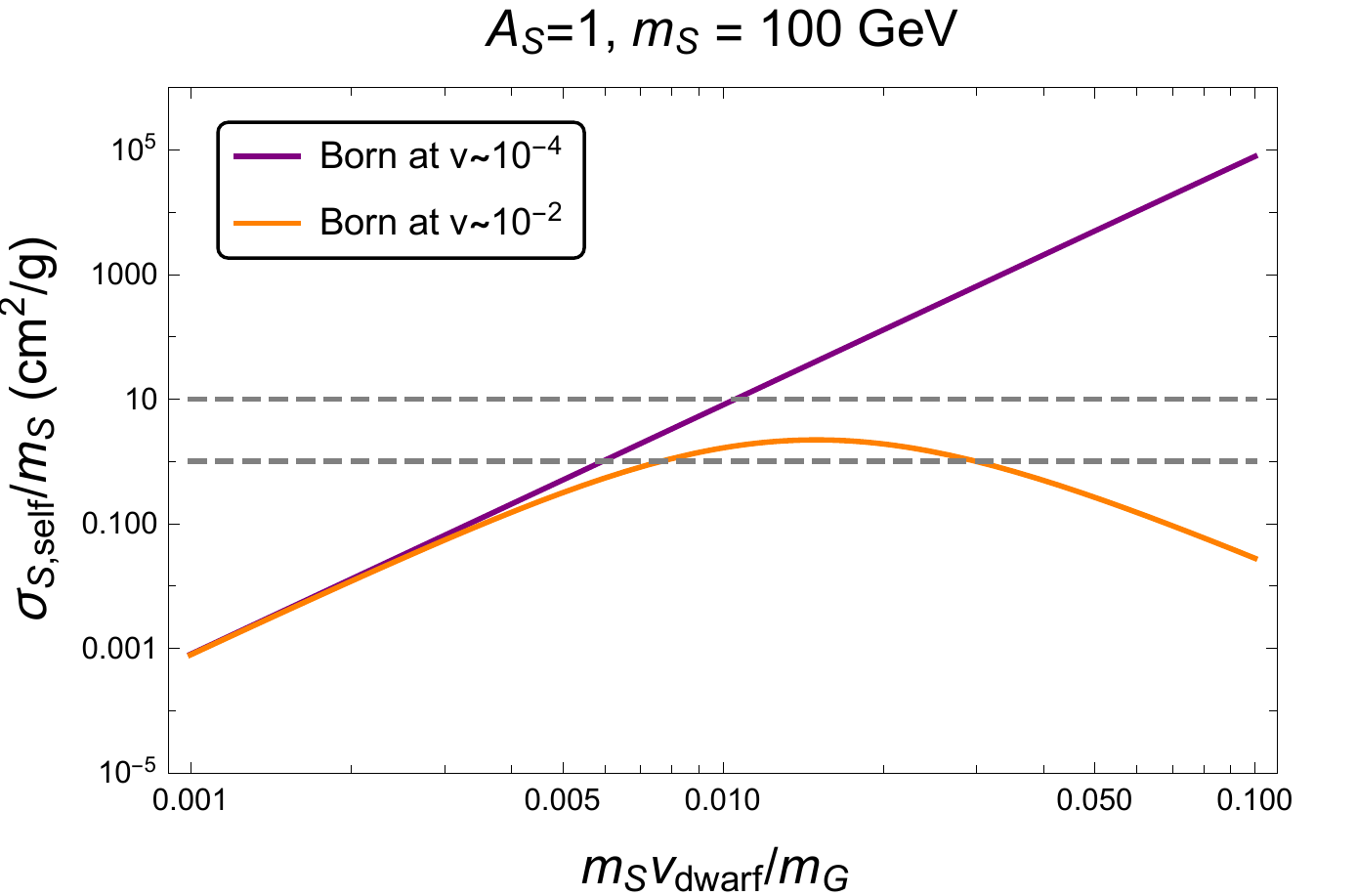}\,\, 
     \includegraphics[height=0.32\textwidth]{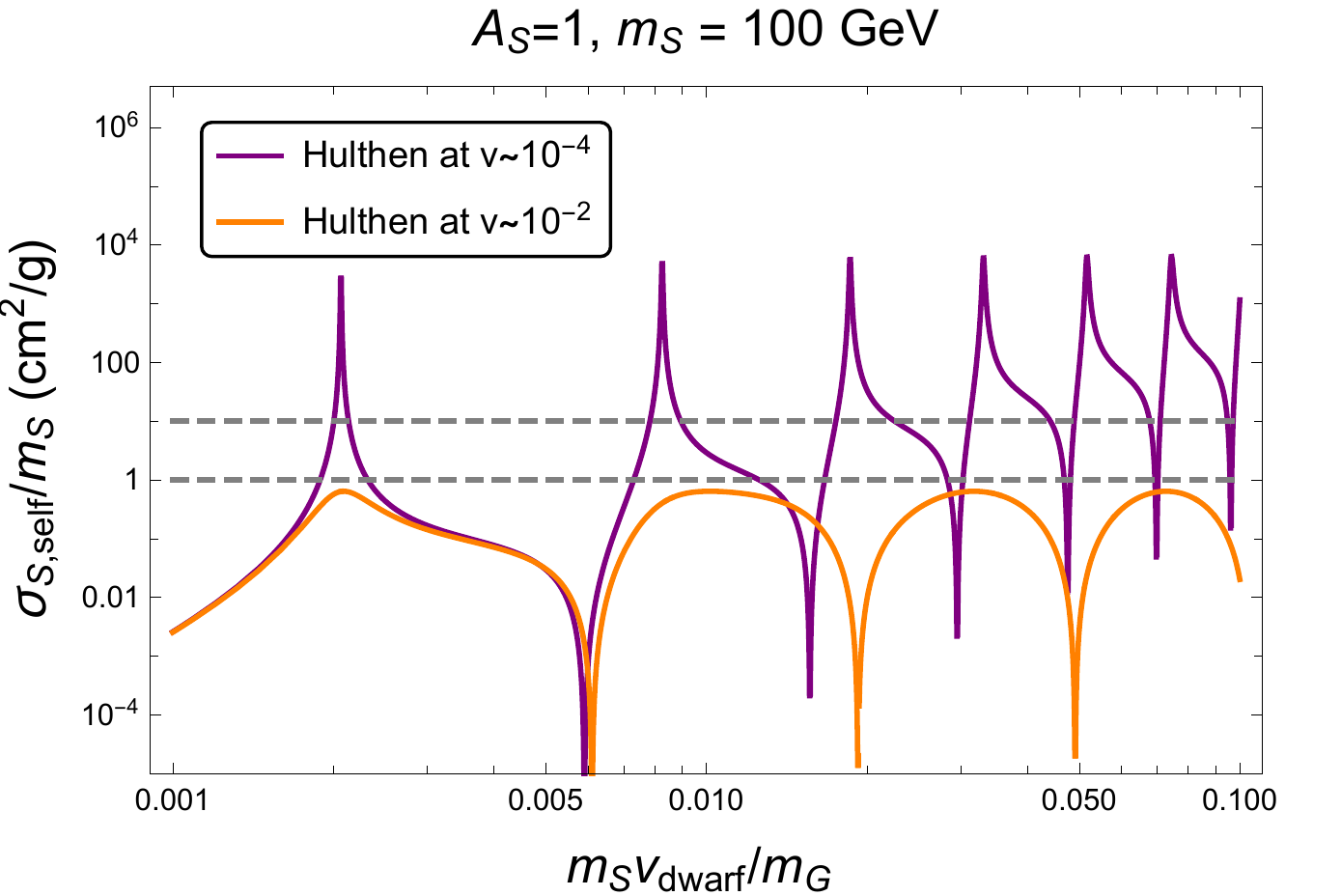}
  \end{center}
  \caption{Born cross section (left) and non-perturbative cross section (right) for the self-scattering of scalar dark matter. We took $A_S=1$ and $m_S=100\,{\rm GeV}$. We made the Hulth\'en potential approximation for the non-perturbative cross section. }
  \label{crossections}
\end{figure}

\begin{figure}[tbp]
  \begin{center}
    \includegraphics[height=0.40\textwidth]{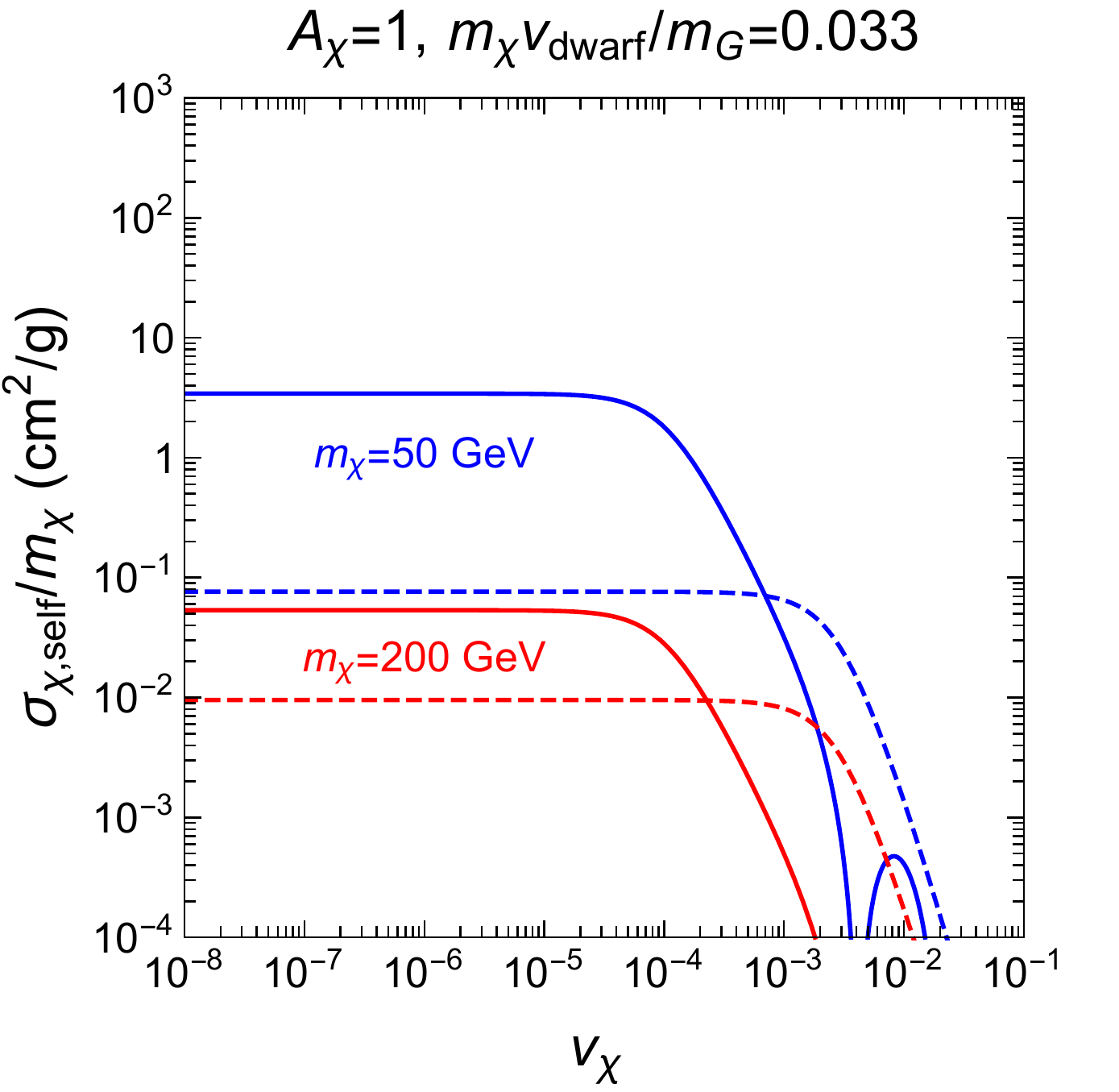}\,\, 
     \includegraphics[height=0.40\textwidth]{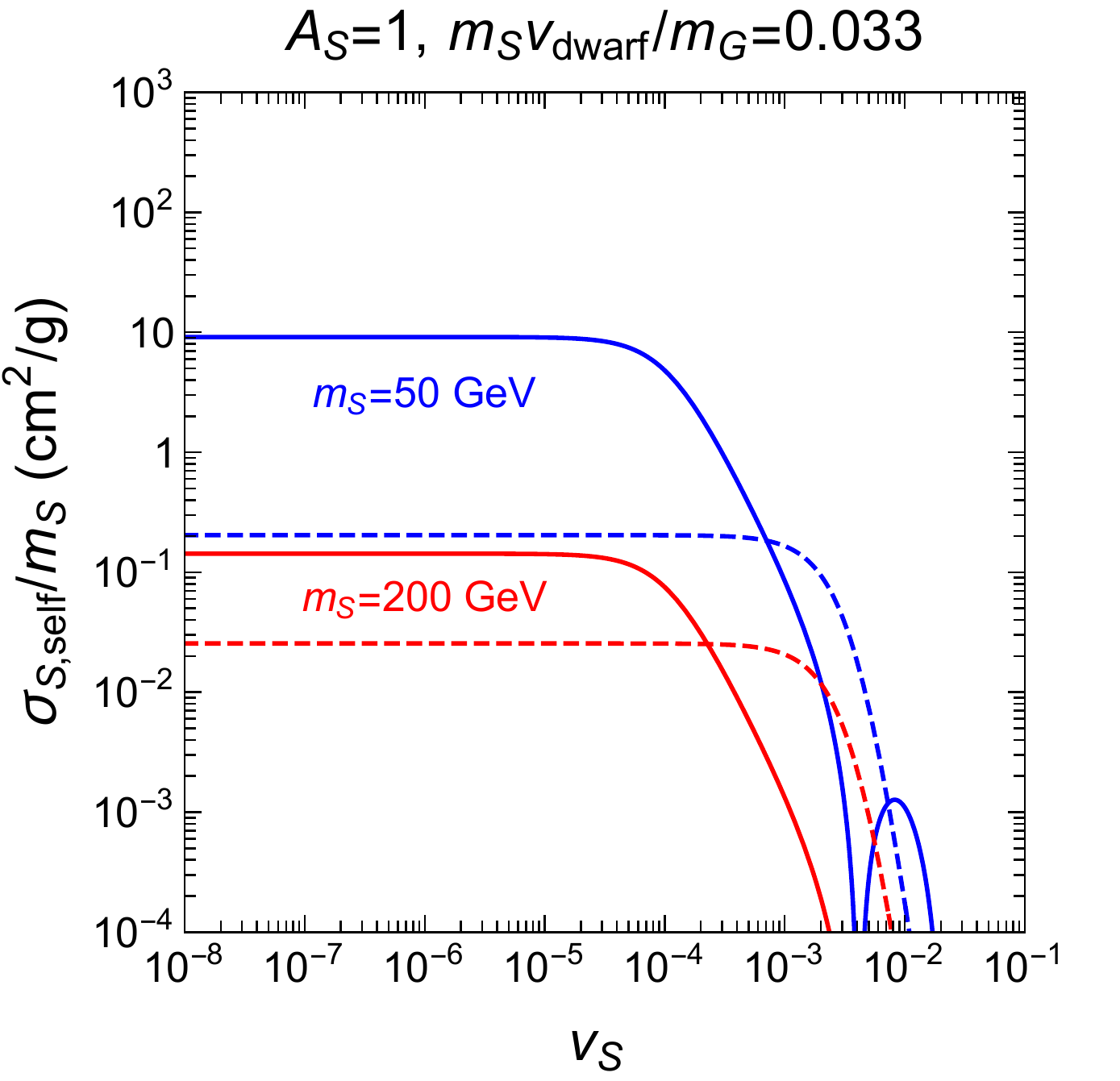}
  \end{center}
  \caption{The DM self-scattering cross section divided by DM mass as a function of the DM velocity, depending on the spins of dark matter, $s=1/2,0$ on left and right. We have chosen  $A_{\rm DM}=1$ and $m_{\rm DM} v_{\rm dwarf}/m_G=0.1$, at dwarf galaxies with $v_{\rm dwarf}=10^{-4}$. Dashed and solid lines are for the Born and non-perturbative cross sections, respectively. The case with $s=1$ shows the similar result as for the case with $s=0$. }
  \label{vs}
\end{figure}

In Fig.~\ref{crossections}, we depict the self-scattering  cross section divided by the DM mass for scalar dark matter as a function of $m_S v_{\rm dwarf}/m_G$. The Born approximation is made on left and the non-perturbative cross section with the Hulth\'en potential approximation is considered on right. We chose the dark matter velocity to $v=10^{-2}$ and $10^{-4}$ in orange and purple lines, and $A_S=1$ and $m_S=100\,{\rm GeV}$ were taken. 
Thus, we can see that the Born cross section is already velocity-dependent and it depends on the mass of the spin-2 mediator. But, there is a clear distinction between the Born and non-perturbative cross sections, due to the resonance effects in the latter case, in particular, at small velocities of dark matter. 
 
On the other hand, in Fig.~\ref{vs}, we also show the DM self-scattering cross section divided by the DM mass as a function of the DM velocity for $A_{\rm DM}=1$ and several choices of the DM and spin-2 mediator masses.
The cases for fermion and scalar dark matter are shown on left and right.  The case for vector dark matter is similar to the case for scalar dark matter, so we don't show it in Fig.~\ref{vs}.
Dashed and solid lines indicate the Born self-scattering cross section and the non-perturbative self-scattering cross section from the Hulth\'en potential, respectively. 
Here, we chose $m_{\rm DM}=50, 200\,{\rm GeV}$ and $m_{\rm DM} v_{\rm dwarf}/m_G=0.033$ (i.e. $m_G=0.15, 0.60\,{\rm GeV}$) for blue and red lines, respectively. In this case, the resulting self-scattering cross section gets saturated to a constant value below $v_{\rm DM}\sim 10^{-4}$ and it becomes highly suppressed at $v_{\rm DM}\sim 10^{-2}$ below the bounds from Bullet cluster \cite{bullet}.
As a result, the self-scattering cross section of dark matter is suppressed at large velocities to be consistent with the disparity between rotation curves of galaxies and galaxy clusters.

We remark that  the velocity dependence of the self-scattering cross section is significant already in the Born limit given in eqs,~(\ref{born1})-(\ref{born3}), so only mild non-perturbative or resonance effects are needed to get sufficiently large values of the self-scattering cross section for WIMP dark matter. Increasing (decreasing)  $A_{\rm DM}$ with the enhancement factor fixed at galaxies, we need to choose a smaller (larger) DM mass or a larger (smaller) spin-2 mediator mass in order to get the enhancement factor suppressed at galaxy clusters.

We note that there is also a possibility to make the self-scattering cross section velocity-dependent by the $s$-channel resonance \cite{murayama}, in the case of vector dark matter of our model, whereas the $s$-channel resonance has an overall velocity-suppression in the cases for scalar or fermion dark matter.

\subsection{Dark matter annihilations}

When dark matter couples to a light spin-2 mediator, it is indispensable for dark matter to  annihilate into a pair of spin-2 mediators, i.e. ${\rm DM\,DM}\rightarrow GG$ is kinematically open and $s$-wave, independent of the spins of dark matter \cite{GMDM}. So, if the mentioned annihilation process dominates in determining the relic density and the spin-2 mediator decays before the CMB recombination, the corresponding annihilation cross section would be enhanced by the Sommerfeld effects at a smaller velocity, thus making the WIMP-like dark matter incompatible with Planck data.  

Adopting the approximate analytic solutions  with the Hulth\'en potential as for the dark matter self-scattering, we obtain the Sommerfeld factor for the $s$-wave dark matter scattering  \cite{cassel}  as
\bea
S_0 = \frac{\frac{\pi}{2}x\,\sinh(2\pi w)}{ \sinh\Big[\pi w\Big(1-\sqrt{1-\frac{x}{w}}\Big)\Big] \sinh\Big[\pi w\Big(1+\sqrt{1-\frac{x}{w}}\Big)\Big]} \nonumber \\
\eea
with $x=\frac{A_{\rm DM}}{4\pi v}$ and $w=\frac{k}{\delta}=\frac{6}{\pi^2} \frac{m_{\rm DM} v}{m_G}$.  Then, the tree-level annihilation cross section $(\sigma v)_0$ for dark matter is replaced by $(\sigma v)_{\rm ann}\simeq S_0\,(\sigma v)_0$, which is enhanced at a low velocity for dark matter. We note that the Sommerfeld factor is saturated to a constant value for $v\lesssim \frac{\pi}{12}\,m_G/m_{\rm DM}$.

Simple solutions to the problem with Sommerfeld-enhanced annihilation cross section for ${\rm DM\,DM}\rightarrow GG$ would be to make the spin-2 mediator long-lived until CMB recombination with small couplings to the SM or make the ${\rm DM\,DM}\rightarrow GG$ annihilation channel subdominant for determining the relic density \cite{hambye} or produce dark matter during the early matter domination \cite{axino}. In the first solution, we could make the spin-2 couplings to the SM small enough and the spin-2 mediator decaying into neutrinos or light particles in the hidden sector \cite{hambye}. In the second solution, there is no need of  a large suppression of the dark matter annihilation into a pair of spin-2 mediators, because we needed relatively mild Sommerfeld effects for velocity-dependent self-interactions. 
If either solutions are not realized, the tree-level cross section for ${\rm DM\,DM}\rightarrow GG$ must be suppressed for satisfying the CMB constraints, thus giving rise to a small self-scattering cross section for dark matter, ${\rm DM\,DM}\rightarrow {\rm DM\,DM}$.

\section{Non-linear interactions and unitarity}

In this section, we discuss the Vainshtein effects on the self-scattering of dark matter and the unitarity bounds on the annihilation of dark matter in massive gravity. 
These effects are distinguishable from the cases with other spins of mediators such as scalar or vector particles for self-interacting dark matter.

\subsection{Vainshtein effects on self-scattering}

There appears a helicity-0 mode in massive gravity at the non-linear level in addition to the five physical degrees of freedom \cite{dRGT,dRGT-review}. In the decoupling limit of massive gravity, the helicity-0 mode $\pi$ can be described by the cubic Galileon theory with the coupling to dark matter \cite{dRGT,dRGT-review}, as follows,
\bea
{\cal L}_{G} = \frac{1}{2} (\partial\pi)^2 -\frac{1}{\Lambda^3_3}\, (\partial\pi)^2 \Box\pi -\frac{c_{\rm DM}}{\Lambda}\, \pi \,T_{\rm DM}
\eea
with $T_{\rm DM}=T^{\rm DM,\mu}_\mu$ and $\Lambda_3=(m^2_G\Lambda/c_{\rm DM})^{1/3}$ is the strong coupling scale in massive gravity.
Then, the helicity-0 mode contribution to the dark matter potential is subject to the Vainshtein effect below the Vainshtein radius $r_*$ in our model, which is given \cite{dRGT,dRGT-review}  by
\bea
r_*& =& \frac{1}{\Lambda_3} \Big(\frac{c_{\rm DM} m_{\rm DM}}{4\pi \Lambda} \Big)^{1/3} \nonumber \\
&=&m_G^{-1}\, \Big(\frac{m_G}{m_{\rm DM}} \Big)^{1/3} \Big( \frac{3A_{\rm DM}}{8\pi}\Big)^{1/3}
\eea
where we used the effective fine-structure constant in eq.~(\ref{fine}) in the second equality.

First, for $m_G\gg m_{\rm DM}$ and $A_{\rm DM}\sim 1$, the Vainshtein radius is much larger than the the range of the Yukawa potential,  $r_G=m^{-1}_G$, so the resulting potential for dark matter due to  the helicity-0 mode would be suppressed by $(r/r_*)^{3/2}$  or $(r/r_*)^2$, depending on the realization of massive gravity theories, in the region with $r\lesssim r_G\ll r_*$ where the Coulomb limit of the potential in eq.~(\ref{Yuk}) exists. In this case, the helicity-0 mode contribution can be safely ignored in our analysis. In the plots in Fig.~\ref{mg}, the region with $m_G\gg m_{\rm DM}$ (to the right of the blue dashed lines) shows that the DM self-scattering cross section is given by $\sigma_T/m_{\rm DM}\ll 0.1\,{\rm cm^2/g}$.

On the other hand, for $m_G\ll m_{\rm DM}$ and $A_{\rm DM}\sim 1$, we find that $r_*\ll m^{-1}_G$, that is, the Vainshtein radius can be much smaller than the range of the Yukawa potential.   Therefore, there is a range of the radius with $r_*<r\lesssim r_G$ for which the helicity-0 mode contribution becomes Coulomb-like so it can be included to capture non-perturbative effects in the effective theory. On the other hand, for $r<r_*$, the Vainshtein screening suppresses the potential due to the helicity-0 mode, so we can ignore the effects of the helicity-0 mode contribution as compared to the Yukawa potential in eq.~(\ref{Yuk}). 
In Figs.~\ref{mg}, \ref{crossections}  and \ref{vs}, in the parameter space where the non-perturbative effects are significant, that is, near the resonance condition given in eq.~(\ref{resonance}),  the Vainshtein radius $r_*$ becomes $r_*\sim m^{-1}_G (3 A_{\rm DM}/(4\pi^2 n))^{2/3}\ll r_G$ for $A_{\rm DM}\sim 1$. In this case, for $r\gtrsim r_*$, the extra contribution of the helicity-0 mode for the dark matter potential is given by 
\bea
\Delta V_{\rm eff}\simeq -\frac{c^2_{\rm DM} m^2_{\rm DM}}{4\pi \Lambda^2 r}, 
\eea
thus leading to a change in the effective self-coupling for dark matter from $A_{\rm DM}$ to $\frac{5}{2}A_{\rm DM}$ for $r_*<r\lesssim r_G$. Therefore, in this case, there is a caution of the interpretation of our results in the previous section, as  $\frac{5}{2}A_{\rm DM}=1$ should be  taken in the plots near the resonance regions in Figs.~\ref{mg}, \ref{crossections} and \ref{vs}. It is interesting to see how the effective potential for dark matter could be affected by the non-linear interactions of the massive spin-2 particle in a concrete realization of ghost-free massive gravity.

\subsection{Unitarity bounds}

The perturbative unitarity is the issue for the dark matter annihilation into a pair of spin-2 mediators. The unitarity scale depends on other couplings of the spin-2 mediators such as quadratic couplings to dark matter and cubic self-couplings \cite{dRGT,unitarity}, without affecting our previous discussion on the DM self-scattering. In particular, non-linear interactions for the massive spin-2 particle are important for the ghost-free realization of a massive spin-2 particle \cite{dRGT,dRGT-review}.

For instance, fixing the quadratic coupling to dark matter and cubic self-couplings for the massive spin-2 mediator appropriately in the dRGT gravity \cite{dRGT}, the unitarity for ${\rm DM}\,G\rightarrow {\rm DM}\,G$ or ${\rm DM\,DM}\rightarrow GG$ by crossing symmetry can be preserved best until the energy scale \cite{unitarity}, given by
\bea
E_{\rm max}\sim \bigg(\frac{m_G\Lambda^2}{c^2_{\rm DM}}\bigg)^{1/3}=\bigg(\frac{2m_G m^2_{\rm DM}}{3A_{\rm DM}}\bigg)^{1/3}.  \label{unitarity}
\eea
Thus, close to the resonance condition for non-perturbative self-scattering or Sommerfeld effects in eq.~(\ref{resonance}), we find that the maximum energy scales for dark matter annihilation processes become $E_{\rm max}\sim \frac{1}{\pi n^{2/3}}\, m_{\rm DM}$, which is independent of the effective fine-structure constant $A_{\rm DM}$ for the spin-2 mediator.

Now we discuss the unitarity scale from non-linear interactions and its effects on the dark matter self-scattering process. 
It is known that the unitarity of the spin-2 mediator self-scattering \cite{unitarity,GLDM} would be violated at  $\Lambda_3= (m^2_G\Lambda/c_{\rm DM})^{1/3}=(m^2_G m_{\rm DM})^{1/3}(3A_{\rm DM}/2)^{-1/6}$, which is parametrically smaller that the one read from ${\rm DM\,DM}\rightarrow GG$.  
Although the massive gravity theory would enter the strong coupling regime at  $\Lambda_3$, the strong coupling scale enters only in the loop processes for the dark matter self-scattering.
The helicity-0 mode $\pi$ could contribute to the dark matter self-scattering at loops, due to the self-interactions, $\frac{1}{\Lambda^3_3}(\partial\pi)^2 \Box \pi$, written in the decoupling limit \cite{dRGT,dRGT-review}, and its linear coupling to dark matter. In this case, as discussed just above, the strong coupling scale $\Lambda_3$ can be smaller than the unitarity scale inferred from the dark matter annihilation, so the loop corrections due to the helicity-0 mode in each ladder diagram for the dark matter self-scattering scale by the factor, $\frac{1}{16\pi^2} \frac{m^4_{\rm DM}}{\Lambda\Lambda^3_3}\sim \frac{1}{16\pi^2}\frac{m^4_{\rm DM}}{\Lambda^2 m^2_G} \sim \frac{1}{16\pi^2} \frac{A_{\rm DM}m^2_{\rm DM}}{m^2_G}$. Therefore, the mass of the spin-2 mediator would be bounded to $m_G\gtrsim \sqrt{(3A_{\rm DM}/2)} \, m_{\rm DM}/(4\pi)$, thus the case with a light spin-2 mediator would be beyond the perturbativity regime.

However, suppose that the unitarity associated with the self-interactions of the massive spin-2 mediator could be ensured by another dynamics to a higher scale such as eq.~(\ref{unitarity}). Then, after replacing $\Lambda_3$ by $E_{\rm max}$ in eq.~(\ref{unitarity}), the loop corrections due to the helicity-0 mode in each ladder diagram for the dark matter self-scattering scale by the factor, $\frac{1}{16\pi^2} \frac{m^4_{\rm DM}}{\Lambda E^3_{\rm max}}\sim \frac{1}{16\pi^2}\frac{m^4_{\rm DM}}{\Lambda^3 m_G} \sim \frac{1}{16\pi^2} (\frac{3}{2})^{\frac{3}{2}}\frac{A^{3/2}_{\rm DM}m_{\rm DM}}{m_G}$. As a result, imposing the resonance condition in eq.~(\ref{resonance}), we get the bound on the effective self-coupling as $A^{1/2}_{\rm DM}\lesssim \frac{16\sqrt{2/3}}{\pi n^2}$ from the perturbativity. 
Under this assumption, the perturbativity for the dark matter self-scattering could be well defined and the effective field theory for the massive spin-2 mediator could be ensured at least in the regimes where the velocity-dependent self-scattering for WIMP dark matter are relevant at galaxies and galaxy cluster scales and the corresponding freeze-out process is taken into consideration.

\section{Conclusions}
We investigated a novel possibility that self-interacting dark matter is endowed to be velocity-dependent due to the exchange of a massive spin-2 particle between dark matter particles. We showed that both the Born self-scattering cross section and the relatively mild non-perturbative effects assist to make the self-interacting cross section velocity-dependent to be compatible with rotation curves of both galaxies and galaxy clusters. Self-interacting dark matter necessarily annihilates into a pair of spin-2 mediators, but the potential problem for CMB recombination can be avoided if there exist other DM annihilation channels or the spin-2 mediator is sufficiently long-lived without visible decay modes. 

We also showed that the Vainshtein effects due the helicity-0 mode could modify the self-scattering of dark matter in the decoupling limit of massive gravity if the Vainshtein radius is smaller than the range of the Yukawa potential between dark matter particles. 
Thus, the massive spin-2 mediator augmented with the helicity-0 mode could make a distinct feature from the case with other typos of mediators such as scalar or vector mediators.
We also found that our model for self-interacting dark matter can be marginally consistent with perturbative unitarity in the ghost-free realization of the massive spin-2 particle.

\section*{Acknowledgments}
We would like to thank Veronica Sanz and Myeonghun Park for discussion at the early stage of the project.
The work is supported in part by Basic Science Research Program through the National Research Foundation of Korea (NRF) funded by the Ministry of Education, Science and Technology (NRF-2019R1A2C2003738 and NRF-2018R1A4A1025334). 
The work of YJK is supported in part by the National Research Foundation of Korea (NRF-2019-Global Ph.D. Fellowship Program).

\appendices%

\section{One-loop corrections to the self-scattering amplitude for dark matter\label{app:loops}}

\begin{figure}[tbp]
\centering
\includegraphics[width=.40\textwidth]{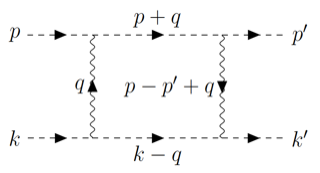}\,\,\,\,
\includegraphics[width=.40\textwidth]{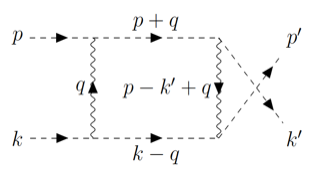}
  \caption{One-loop Feynman diagrams for the self-scattering of scalar dark matter.  }
  \label{loops}
\end{figure}

We discuss the details of the one-loop corrections to the self-scattering amplitude for scalar dark matter. We have shown the tree-level amplitude in Section 3.2. In comparison, from the Feynman diagrams with two spin-2 particle exchanges in Fig.~\ref{loops}, we obtain the one-loop $t$-channel amplitude for the self-scattering of two scalar dark matter particles, $S(p)+S(k)\rightarrow S(p')+S(k')$, as follows,
\bea
i\Gamma^t_{\rm loop} &=&\Big(\frac{c_S}{\Lambda}\Big)^4 \int \frac{d^4 q}{(2\pi)^4} \frac{1}{(q^2-m^2_G) ((q+p)^2-m^2_S)((q-k)^2-m^2_S)((p-p'+q)^2-m^2_G)} \nonumber \\
&&\times \Big(\tau_{\mu\nu}(p,q+p) \tau_{\alpha\beta}(k,k-q) P^{\mu\nu,\alpha\beta}(q) \Big)  \nonumber \\
&&\times  \Big(\tau_{\rho\sigma}(q+p,p') \tau_{\kappa\epsilon}(k-q,k') P^{\rho\sigma,\kappa\epsilon}(p-p'+q) \Big) 
\eea
where
\bea
\tau_{\mu\nu}(k,q) = 2 k_\mu q_\nu + (m^2_S- k\cdot q) \eta_{\mu\nu}.
\eea
Here, we have used $P^{\mu\nu,\alpha\beta}=P^{\nu\mu,\alpha\beta}=P^{\mu\nu,\beta\alpha}$ in writing the energy-momentum tensor in the above form $\tau_{\mu\nu}$. 
For the $u$-channel one-loop diagram, we can obtain the corresponding amplitude from interchanging $p\leftrightarrow k$ in $i\Gamma^t_{\rm loop}$.

Then, using the Feynman parameters, 
\bea
\frac{1}{a^2_1 a_2 a_3}= 3! \int^1_0 dx \int^{1-x}_0 dy \frac{1-x-y}{\big(a_1(1-x-y)+a_2 y+ a_3 x )^4},
\eea
we can rewrite the one-loop scattering in the limit of the forward scattering with $p=p'$ as
\bea
i\Gamma^t_{\rm loop} &=& 3! \Big(\frac{c_S}{\Lambda}\Big)^4 \int^1_0 dx \int^{1-x}_0 dy\,  (1-x-y) \,  \int \frac{d^4q}{(2\pi)^4} \frac{
1}{({\tilde q}^2-\Delta^2)^4} \nonumber \\
&&\times \Big(\tau_{\mu\nu}(p,q+p) \tau_{\alpha\beta}(k,k-q) P^{\mu\nu,\alpha\beta}(q) \Big)  \nonumber \\
&&\times  \Big(\tau_{\rho\sigma}(q+p,p') \tau_{\kappa\epsilon}(k-q,k') P^{\rho\sigma,\kappa\epsilon}(p-p'+q) \Big) 
\eea
with ${\tilde q}=q+y p- x k$. 
As a result, for the non-relativistic and forward scattering of scalar dark matter, we get the approximate results for the one-loop amplitude in dimensional regularization with $d=4-2\epsilon$, as follows,
\bea
i\Gamma^t_{\rm loop} &=&  \frac{c_S^4}{1728\pi^2 m_G^4 m_S^2 \Lambda^4} \bigg[ 24 m_G^8 
    m_S^2 - 222 m_G^6 m_S^4 + 
        353 m_G^4 m_S^6 + 52 m_G^2 m_S^8 - 12 m_S^{10}  \nonumber \\
        &&- 
        3 (4 m_G^{10} - 43 m_G^8 m_S^2 + 146 m_G^6 m_S^4 - 130 m_G^4 m_S^6 - 
      100 m_G^2 m_S^8 - 
             168 m_S^{10}) \ln \Big(\frac{m_G^2}{m_S^2} \Big) \nonumber \\
             &&+ 
   6(m_G^2 - 4m_S^2)^2 \sqrt{m_G^4-4 m_G^2 m_S^2 }
          \nonumber \\
          &&\quad\times (4 m_G^4 - 3 m_G^2 m_S^2 - 4 m_S^4)
    \ln\bigg(\frac{m_G^2 + \sqrt{m_G^4 - 4 m_G^2 m_S^2 }}{
              2 m_G m_S}\bigg)  \nonumber \\
              &&+ 
   72 m_S^8 (10m_G^2 + 7m_S^2)\bigg(\frac{1}{\epsilon}+\ln\Big(\frac{\mu^2}{m_G^2}\Big) \bigg) \bigg]. \label{fulloneloop}
\eea
Therefore, the one-loop amplitude becomes divergent due to $1/\epsilon$ in the last line, but the divergent part is cancelled by the renormalization of the quartic coupling for scalar dark matter. We note that the $u$-channel diagram leads to the same result, $\Gamma^u_{\rm loop}\approx\Gamma^t_{\rm loop} $.

On the other hand, for $\xi\equiv m^2_S/m^2_G\gg 1$, we get the approximate result for the finite part of the one-loop amplitude from eq.~(\ref{fulloneloop}), as follows,
\bea
i\Gamma^t_{\rm loop,finite}  \approx \frac{2c^4_S m^6_S}{9\pi \Lambda^4 m^2_G}\,\sqrt{\xi}.
\eea

We also remark on the velocity-dependent loop corrections to the self-scattering amplitude. 
For the non-relativistic forward scattering, there are leading velocity-dependent contributions to the one-loop amplitude, given by
\bea
i\delta \Gamma^t_{{\rm loop},v}\approx \frac{c_S^4m^6_S }{216 \pi^2 m^4_G \Lambda^4}\, (660m^2_G+211 m_S^2)\, \cdot\frac{v^2}{\epsilon} -\frac{8 c^4_S m^4_S}{9\pi \Lambda^4}\, v^2 \xi^{5/2}
\eea
where $v$ is the relative velocity between two dark matter particles.
Therefore, we also need to introduce higher dimensional  counter terms for scalar dark matter,  such as $c_6(\partial_\mu S\partial^\mu S) S^2$, etc, to cancel the velocity-dependent divergent terms. On the other hand, the finite velocity-dependent loop corrections can be ignored as compared to the velocity-independent corrections, as far as $4 v^2\xi\lesssim 1$, that is, for a sufficiently small velocity of dark matter. This is true of dark matter in galaxies and galaxy clusters that we are interested in.

\end{document}